\documentclass[aps,prl,reprint,showpacs,superscriptaddress]{revtex4-1}

\usepackage{tikz}
\usepackage{pgfplots}
\usepackage{caption}
\usepackage{float}
\usetikzlibrary{plotmarks}
\usetikzlibrary{snakes,arrows,decorations.pathmorphing,backgrounds,positioning,fit,petri}

\usetikzlibrary{calc} 

\usepackage{hyperref}
\hypersetup{
	bookmarksopen=true,%
	bookmarksnumbered=true,%
	colorlinks=true,%
    	linkcolor=blue,
    	citecolor=blue,
    	filecolor=blue,
    	urlcolor=blue,
	pdfstartview=FitH,%
	pdfnewwindow=true
}

\usepackage{graphicx}

\usepackage{mathtools} \mathtoolsset{showonlyrefs}

\usepackage{txfonts}

\usepackage{microtype}

\newcommand{\bra}[1]{\langle#1\rvert}
\newcommand{\ket}[1]{\lvert#1\rangle}

\makeatletter
\newcommand{\vast}{\bBigg@{2}}
\newcommand{\Vast}{\bBigg@{4}}
\makeatother

\newcommand{\mi}{\mathrm{i}}

\begin{document}
\title{Stimulated emission from a single excited atom in a waveguide}
\author{Eden Rephaeli}
\email{edenr@stanford.edu}
\affiliation{Department of Applied Physics, Stanford University, Stanford, California 94305, USA}
\author{Shanhui Fan}
\email{shanhui@stanford.edu}
\affiliation{Department of Electrical Engineering, Stanford University, Stanford, California 94305, USA}

\date{\today}

\begin{abstract}
We study stimulated emission from an excited two-level atom coupled to a waveguide containing an incident single-photon pulse. We show that the strong photon correlation, as induced by the atom, plays a very important role in stimulated emission.  Additionally, the temporal duration of the incident photon pulse is shown to have a marked effect on stimulated emission and atomic lifetime.
\end{abstract}
\maketitle


\pdfbookmark[1]{Introduction}{Introduction}
\paragraph{Introduction}   
Stimulated emission, first formulated by Einstein in 1917 \cite{Einstein1917}, is the fundamental physical mechanism underlying the operation of lasers \cite{Siegman1986} and optical amplifiers \cite{Lamas-Linares2002,Sun2007}, both of which are of paramount importance in modern technology. In recent years, stimulated emission has been studied in a variety of novel systems, including a surface plasmon nanosystem \cite{Bergman2003}, a single molecule transistor \cite{Hwang2009}, and superconducting transmission lines \cite{Astafiev2010}. Stimulated emission also plays a crucial role in the quantum cloning of photons \cite{Simon2000,Lamas-Linares2001,Lamas-Linares2002,Fasel2002}. 

In textbooks \cite{Saleh1991}, stimulated emission is usually described by having a photon interact with an atom in the excited state. As a result of such interaction, the atom emits a second photon that is ``identical" to the incident photon. As a concrete experimental example, in Refs. \cite{Sun2007,Liu2009}, photons of a given polarization were sent into a parametric amplifier, and stimulated emission manifested in an enhanced probability of the outgoing photons having the same polarization.  In Refs. \cite{Sun2007,Liu2009}, such enhancement was attributed entirely to constructive interference due to photon indistinguishability.

\begin{figure}
\includegraphics[scale=0.7]{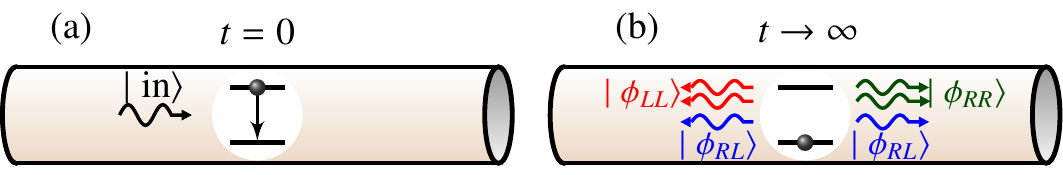}
\caption{(Color online) Schematic of the atom-waveguide system studied here. (a) The initial state at t=0, consisting of an incident single-photon pulse, and an atom in the excited state. (b) At $t\to\infty$ the atom is in the ground-state, and the photon field is in a superposition of two-photon states $\mid \phi_{RR}\rangle$,$\mid \phi_{LL}\rangle$, and $\mid \phi_{RL}\rangle$ (green, red, and blue, respectively).}  
\label{F:1}
\end{figure}

In this Letter, we consider a scenario of stimulated emission that is arguably closest to the textbook description \cite{Saleh1991}. We consider an atom coupled to a waveguide, and study the interaction of the atom with a single incident photon, when the atom is initially in the excited state, as shown in  Fig. \ref{F:1}(a). The atom-waveguide system is of current experimental and theoretical interest both for nanophotonics in the optical regime \cite{Shen2005,Shen2007,Shen2007a,Yudson2008,Zheng2010,Shi2010,Chang2007} and in the studies of superconducting transmission lines in the microwave regime \cite{Wallraff2004,Houck2007}. In an atom-waveguide system, the ability to amplify few-photon quantum states is important, and has been recently demonstrated experimentally \cite{Astafiev2010}. Thus, the scenario that we consider is of experimental interest. 

We observe that the Hamiltonian describing the waveguide-atom system also describes, in 3D, the interaction of a light beam with an atom, when the beam is designed to mode-match the atom's radiation pattern.  This observation was emphasized and exploited in a number of recent papers \cite{Gerhardt2007,Zumofen2008,Hwang2009,Hetet2011}.  Thus, our result should be relevant to a wide class of recent 3D experiments as well.

In recent years, there has been much advancement in the capacity to deterministically generate single photons \cite{Kuhn2002,McKeever2004} and to control the shape of the single-photon pulse \cite{Kolchin2008, Specht2009}.  In both waveguide and free-space, a single photon, by necessity, must exist as a pulse \cite{Chan2002}. Therefore, our study of stimulated emission at the single-photon level reveals important dynamic characteristics of stimulated emission.  For instance, we will show that there exists an optimal spectral bandwidth of the single-photon pulse that results in the largest enhancement of forward scattered light.  This is fundamentally different from the conventional study of stimulated emission involving a single-mode continuous radiation field.  Finally, our results show that in the waveguide-atom system, emission enhancement in the stimulated process cannot be entirely attributed to photon indistinguishability but, instead, has a substantial connection to the strong photon correlation induced by the two-level atom. 

\pdfbookmark[2]{Model system and the incident single-photon pulse}{Model system and the incident single-photon pulse}
\paragraph{Model system and the incident single-photon pulse}
We consider the atom-waveguide system shown in Fig. \ref{F:1}. 
For a waveguide with a linear dispersion relation, the real-space Hamiltonian of the system in the rotating-wave approximation is \cite{Shen2005}
\begin{align}
&\hat{H}/\hbar=-\mi v_g\int dx c^{\dag}_R(x)\frac{\partial}{\partial x}c_R(x)+\mi v_g\int dx c^{\dag}_L(x)\frac{\partial}{\partial x}c_L(x)+\frac{\Omega}{2}\sigma_z \\&+\frac{V}{\sqrt{2}}\int dx\delta(x)\left[c^{\dag}_R(x)\sigma_-+c^{\dag}_L(x)\sigma_-+\sigma_+c_R(x)+\sigma_+c_L(x)\right],\label{E:H}
\end{align}
where $x$ is the spatial coordinate along the waveguide's symmetry axis. $c^{\dag}_R(x)\left[c^{\dag}_L(x)\right]$ creates a right [left] moving photon,  $c_R(x)\left[c_L(x)\right]$ annihilates a right [left] moving photon, and $\sigma_+(\sigma_-)$ is a raising (lowering) operator of the two-level atom.  The atom-waveguide coupling strength is $V$, corresponding to an atom spontaneous emission rate $\Gamma\equiv V^2/v_g$.  The transition frequency of the atom is $\Omega$. $v_g$ is the waveguide group velocity, which we set as $v_g=1$ from here on out.

At $t=0$ , we assume that the atom is in the excited state. At the same time, a right-going single-photon pulse starts to interact with the atom. The in state of the system is therefore
\begin{align}
\ket{\text{in}}=\int dx\psi(x)c^{\dag}_R(x)\sigma_+\ket{0}.
\label{E:psiin}
\end{align}
At $t\to\infty$ the atom has decayed to the ground state, and the out state contains only two-photon states, as given by 
\begin{align}
\ket{\text{out}}=\ket{\phi_{\text{RR}}}+\ket{\phi_{\text{LL}}}+\ket{\phi_{\text{RL}}},
\label{E:psifinal}
\end{align}
where e.g., $\ket{\phi_{\text{RR}}}=\int dx_1dx_2\phi_{RR}(x_1,x_2)(1/\sqrt{2})c^{\dag}_R(x_1)c^{\dag}_R(x_2)\ket{0}$ denotes an outgoing two-photon state with two right-moving photons and $P_{RR}\equiv\ \langle \phi_{RR}\ket{\phi_{RR}}$ is its respective probability.  Here, we use $\phi$ to denote all outgoing states and use subscripts to distinguish these various states. Since the incident photon is right-moving, to study stimulated emission we will be particularly interested in comparing behaviors of $P_{RR}$, where both outgoing photons are right-moving, to other outcomes as described by $P_{RL}$ and $P_{LL}$.

For computation in this Letter we choose an input wavefunction
\begin{align}
\psi(x)=i\sqrt{\alpha\Gamma}e^{i\left(\Omega-i\alpha\Gamma/2\right)x}\theta(-x).\label{E:pulse}
\end{align}
For $\alpha=1$, $\psi(x)=\phi_{sp}(x)$, where $\phi_{sp}(x)$ is the wavefunction of a spontaneously emitted photon by the excited atom.  Variation of $\alpha$ allows us to probe various time and frequency scales, as well as the connection between spontaneous and stimulated emission in radiation dynamics.
\begin{figure}
\includegraphics[scale=1]{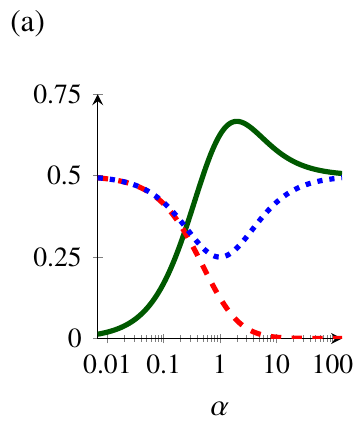}
\includegraphics[scale=1]{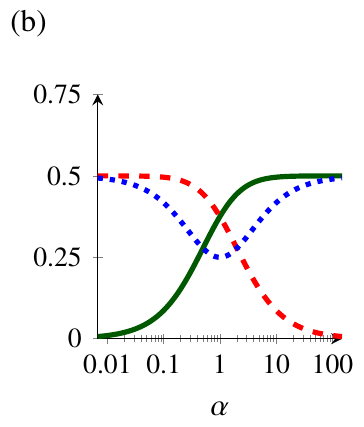}
\includegraphics[scale=1]{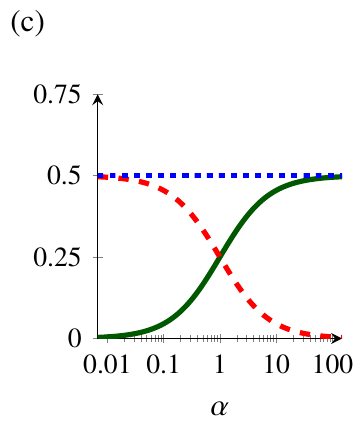}
\caption{(Color online) (a) $P_{\text{RR}},P_{\text{LL}},P_{\text{RL}}$ in solid green, dashed red, and dotted blue lines for a two-level atom.  (b) $P^{cavity}_{\text{RR}},P^{cavity}_{\text{LL}},P^{cavity}_{\text{RL}}$ in solid green, dashed red, and dotted blue lines. (c) $P^{classical}_{\text{RR}},P^{classical}_{\text{LL}},P^{classical}_{\text{RL}}$ in solid green, dashed red, and dotted blue lines.}
  \label{F:2}
\end{figure}

\pdfbookmark[3]{The out state}{The out state}
\paragraph{The out state.}
To calculate the out state for an arbitrary initial state as described by Eq. \eqref{E:psiin}, it is useful to define $c^{\dag}_e(x)\equiv (1/\sqrt{2})\left[c^{\dag}_R(x)+c^{\dag}_L(-x)\right]$ and $c^{\dag}_o(x)\equiv (1/\sqrt{2})\left[c^{\dag}_R(x)-c^{\dag}_L(-x)\right]$.  With these definitions, the Hilbert space is separated into even and odd subspaces, each described by a Hamiltonian:
\begin{align}
&\hat{H}_e/\hbar=-i\int dx c^{\dag}_e(x)\frac{\partial}{\partial x}c_e(x)\\&+V\int dx\delta(x)\left[c^{\dag}_e(x)\sigma_-+\sigma_+c_e(x)\right]+\frac{\Omega }{2}\sigma_z,\label{E:He}\\&\hat{H}_o/\hbar=-i\int dx c^{\dag}_o(x)\frac{\partial}{\partial x} c_o(x).
\end{align} 
We note that $H=H_e+H_o$ and $\left[H_{e},H_{o}\right]=0$.  Similarly, we decompose the in-state:
\begin{align}
&\ket{\text{in}}=\ket{\text{in}}_{eo}+\ket{\text{in}}_{ee}=\frac{1}{\sqrt{2}}\int dx_1dx_2\psi(x)c^{\dag}_o(x)\sigma_+\ket{0}
\\&+\frac{1}{\sqrt{2}}\int dx_1dx_2\psi(x)c^{\dag}_e(x)\sigma_+\ket{0}.
\end{align}
We note that, since $\left[\sigma_+,\hat{H}_o\right]=0$, the state $\sigma_+\ket{0}$ is contained in the even subspace.

We can now calculate the out state by evaluating $\lim_{t\to\infty} e^{-iHt}\ket{\text{in}}$.  From this point on, with the variable $t$ we always assume that we are in the $t\to\infty$ limit.  We consider $\ket{\text{in}}_{eo}$ and $\ket{\text{in}}_{ee}$ separately:
\begin{align}
&e^{-iHt}\ket{{\text{in}}}_{eo}=
\\&=\frac{1}{\sqrt{2}}\iint dx_1dx_2\phi_{\text{sp}}(x_1-t)\psi(x_2-t)\ket{x_1,x_2}_{eo}
\\&\equiv\frac{1}{\sqrt{2}}\iint dx_1dx_2\phi_{eo}(x_c-t,x_d)\ket{x_1,x_2}_{eo},
\label{E:psiouteo}
\end{align}
where  $x_c\equiv \frac{x_1+x_2}{2}$, $x_d\equiv x_1-x_2$, and $\ket{x_1,x_2}_{eo}=c^{\dag}_e(x_1)c^{\dag}_o(x_2)\ket{0}$.
Similarly,
\begin{align}
&e^{-iHt}\ket{\text{in}}_{ee}=
\\&=\frac{1}{\sqrt{2}}\iint dk_1dk_2\int dxe^{-iH_et}\ket{k_1k^+_2}\bra{k_2^+k_1}\psi(x)c^{\dag}_e(x)a^{\dag}\ket{0}
\\&=\iint dx_1dx_2\phi_{ee}(x_c-t,x_d)\ket{x_1,x_2}_{ee},\label{E:psioutee}
\end{align}
where $\ket{k_1k_2^+}$ is the two-excitation interacting eigenstate of $\hat{H}_e$ in the two-excitation manifold, as determined in Ref. \cite{Shen2007a}, and $\ket{x_1,x_2}_{ee}=(1/\sqrt{2})c^{\dag}_e(x_1)c^{\dag}_e(x_2)\ket{0}$.

For the specific initial state as defined in Eq. \eqref{E:psiin}, $\phi_{ee}(x_c-t,x_d)$ and $\phi_{eo}(x_c-t,x_d)$ are calculated in the Supplementary Information below, and as a result we have
\begin{align}
&P_{RR,LL}(\alpha)=\frac{1}{4}\left[1+\frac{2\alpha}{\left(1+\alpha\right)^2}\pm\frac{\left(1+\alpha\right)^3-8\alpha}{(\alpha-1)(1+\alpha)^2}\right],
\\&P_{RL}(\alpha)=\frac{1}{2}\left[1-\frac{2\alpha}{\left(1+\alpha\right)^2}\right],\label{E:probs}
\end{align}
which we plot in Fig. \ref{F:2}(a).
\pdfbookmark[4]{Analysis of Stimulated Emission}{Analysis of Stimulated Emission}
\paragraph{Analysis of Stimulated Emission}
When $\alpha<<1$, $P_{RR}(\alpha)\to 0$ and $P_{RL}(\alpha),P_{LL}(\alpha) \to 0.5$.  In this limit, the temporal duration of the incident photon pulse is much longer than the spontaneous emission lifetime.  Thus, the initial excitation of the atom decays spontaneously, and the pulse is completely reflected by an atom largely in the ground state \cite{Shen2005}.  When $\alpha>>1$, $P_{RR}(\alpha), P_{RL}(\alpha)\to 0.5$, and $P_{LL}(\alpha)\to 0$.  In this case, the photon pulse is extremely short; there is therefore no atom-photon interaction.  As a result, the incident photon pulse is fully transmitted, and the atom decays spontaneously after the pulse passes through.  This limit of $\alpha>>1$ corresponds to a scenario where the transmission of the incident photon and the decay of the atom occur independently. In this "independent" scenario, $P_{RR}=0.5$.  Therefore, $P_{RR}>0.5$ indicates an enhanced probability that both photons go to the right and is therefore a direct indication of stimulated emission.

  In Fig. \ref{F:2}(a), we indeed observe $P_{RR}>0.5$ in the intermediate region of $\alpha$.  In fact, $P_{RR}$ maximizes to $2/3$ at $\alpha=2$---when the temporal decay rate of the incident photon pulse is twice the atom's spontaneous emission rate.  Moreover, the enhancement of $P_{RR}$ can be observed with other incident pulse shapes.  For example, numerically, we have found that an incident photon with half-Gaussian wave function $\psi(x)=\left(\frac{2\alpha^2\Gamma^2}{\pi}\right)^{1/4}e^{-(\alpha\Gamma)^2x^2/4}\theta(-x)$  attains a similar maximum of $P_{RR}\approx 0.65$ for $\alpha\approx 1.6$.

To better understand the stimulated emission behavior in Fig. \ref{F:2}(a), we consider the general expression of the three probabilities $P_{RR},P_{LL}$, and $P_{RL}$, for an arbitrary incident wavefunction $\psi(x)$.  From Eqs. \eqref{E:psiouteo}, \eqref{E:psioutee} one can show that 
\begin{align}
&P_{RL}=0.5-0.25\iint dx_1dx_2\ \phi_{eo}^{*}(x_c-t,-x_d)\phi_{eo}(x_c-t,x_d)
\\&=0.5-0.25\left[\int dx \psi^{*}(x)\phi_{\text{sp}}(x)\right]^2,\label{E:Prl}\end{align}
where to obtain the first line, we note that the cross terms between $\phi_{eo}(x_c-t,x_d)$ and $\phi_{ee}(x_c-t,x_d)$ happen to cancel, and in obtaining the second line we have used Eq. \eqref{E:psiouteo}.  We define a \textit{photon indistinguishability factor}
\begin{align}
F\equiv\left[\int dx \psi^{*}(x)\phi_{\text{sp}}(x)\right]^2,\label{E:F}
\end{align}
which measures the indistinguishability between the incident single-photon pulse and the single-photon pulse from spontaneous emission of an excited atom.  When $\alpha=1$, $F$ maximizes at 1.  Consequently, $P_{RL}(\alpha)$ is minimal.  Moreover, since $0\leq F\leq1$, we have $0.25\leq P_{RL}\leq 0.5$, establishing an upper bound of $3/4$ for $P_{RR}$.

Similarly, from Eqs. \eqref{E:psiouteo} and \eqref{E:psioutee}, we can derive 
\begin{align}&P_{RR,LL}=\frac{1}{4}+\frac{1}{8}F\\&\pm\frac{1}{8}\int dx_1dx_2 \ \Bigg\{\left[\phi^{*}_{eo}(x_c-t,x_d)+\phi^{*}_{eo}(x_c-t,-x_d)\right]\phi_{ee}(x_c-t,x_d)\\&+2\phi_{ee}^{*}(x_c-t,x_d)\phi_{eo}(x_c-t,x_d)\Bigg\},\label{E:Prrll}
\end{align}
In Eq. \eqref{E:Prrll}, we see that $P_{RR}$ and $P_{LL}$ (with + and - signs, respectively) are influenced in the same way by the photon indistinguishability factor $F$ .  Thus, the most prominent feature of stimulated emission in terms of $P_{RR}>0.5$ cannot be attributed to photon indistinguishability.

 The last term in Eq. \eqref{E:Prrll} contains overlap integrals between the out state $\phi_{eo}(x_c-t,x_d)$ in the $eo$ subspace---which is a product of the incident wavefunction $\psi(x-t)$ and a spontaneously emitted photon $\phi_{sp}(x-t)$---and the out state $\phi_{ee}(x_c-t,x_d)$ in the $ee$ subspace. Since a single two-level atom cannot simultaneously absorb two photons, the two-photon wave function $\phi_{ee}(x_c-t,x_d)$ in the $ee$ subspace exhibits complex entangled structures in spatial, spectral, and temporal domains \cite{Shen2007a}.  It is known that the two-photon wavefunction in the $ee$ subspace has a direct connection to the Bethe-ansatz wave functions that have been commonly used to describe strongly correlated quantum systems \cite{Shen2007a}.  Hence, here we refer to the effects arising from the complex structure in $\phi_{ee}(x_c-t,x_d)$ as a strong photon correlation effect.  The form of Eq. \eqref{E:Prrll} therefore directly points to the role that the strong correlation induced by the atom plays in stimulated emission in addition to the effect of photon indistinguishability.
 
The role of photon indistinguishability in stimulated emission is a topic of current interest.  In Ref. \cite{Sun2007}, Sun et. al. studied this question by comparing two scenarios.  In the first scenario, an N-photon state is sent into a parametric down-converter and the amplification results in the creation of an N+1 photon state in the output.  In the second scenario, the amplification process is instead emulated by combining the N-photon state with an extra photon at a beam splitter.  The authors then showed experimentally that these two scenarios produce the same photon statistics.  Thus, the crucial aspect of stimulated emission in their system is the indistinguishability between the photons.

Motivated by their work, here we also consider an alternative scenario where we replace the atom with a side-coupled cavity.  We emulate the stimulated emission process by injecting a single-photon pulse into the waveguide while there is a photon in the cavity.  Mathematically, this system as well as the initial state can be described by replacing $\sigma_-\ (\sigma_+)$ with $a\ ( a^{\dag})$ and $\sigma_z$ with $2a^{\dag}a-1$ in Eqs. \eqref{E:H} and \eqref{E:psiin}. Here $a\ (a^{\dag})$ is the annihilation (creation) operator of a cavity photon, and $[a, a^{\dag}] =1$.  Also, we note that the waveguide-atom and waveguide-cavity systems have identical behaviors in the single-excitation manifold. 
In the absence of the injected single photon in the waveguide, the photon in the cavity decays spontaneously.  Thus, photon indistinguishability in this case refers to how close the photon pulse matches the spontaneously decayed photon pulse from the cavity, as characterized by the same $F$ factor in Eq. \eqref{E:F}.  For the waveguide-cavity system, the output probabilities for various outcomes are
\begin{align}
P^{cavity}_{RR,LL}=\frac{1}{4}\left[1+\frac{2\alpha}{(1+\alpha)^2}\pm\frac{\alpha-1}{\alpha+1}\right],
\label{E:Pcavity}
\end{align}
\begin{align}
P^{cavity}_{RL}(\alpha)=P_{RL}(\alpha).
\end{align}
which we plot in Fig. \ref{F:2}(b).

We compare the waveguide-cavity system to a classical case consisting of a waveguide and an ancilla.  We consider an incident single-photon pulse with a probability spectrum $|\psi(k)|^2=\frac{1}{2\pi}\frac{\alpha\Gamma}{\left(k-\Omega\right)^2+\left(\alpha\Gamma/2\right)^2}$, where $\psi(k)$ is the momentum-space representation of $\psi(x)$ in Eq. \eqref{E:pulse}. The ancilla emits a classical photon with probability $1/2$ in each direction and, moreover, reflects an incident photon with a probability spectrum $|r_{k}|^2=\frac{(\Gamma/2)^2}{(k-\Omega)^2+(\Gamma/2)^2}$.  The ancilla therefore has the same intensity response function to a single photon as either an atom or a cavity. For this classical case we have $P^{classical}_{RR}(\alpha)=\frac{1}{2}\frac{\alpha}{1+\alpha}$, $P^{classical}_{LL}(\alpha)=\frac{1}{2}\frac{1}{1+\alpha}$ and $P^{classical}_{RL}(\alpha)=1/2$, as plotted in Fig. \ref{F:2}(c).  We see that the waveguide-cavity system has an enhanced probability $P^{cavity}_{RR}$ compared to the classical system $P^{classical}_{RR}$.  Photon indistinguishability, which is unique to the quantum system, certainly plays a role here.

However, $P^{cavity}_{RR}(\alpha)$ is always smaller than 0.5.  Thus, the excess probability of having both outgoing photons propagating to the right, i.e., the fact that $P_{RR}(\alpha)>0.5$, cannot be achieved in the waveguide-cavity system and, therefore, cannot be attributed to photon indistinguishability.  It follows that the key difference between the atom and the parametric down-converter in Ref. \cite{Sun2007} is the photon correlation induced by the atom due to its inability to absorb more than one photon at a time.

\begin{figure}
\includegraphics[scale=1]{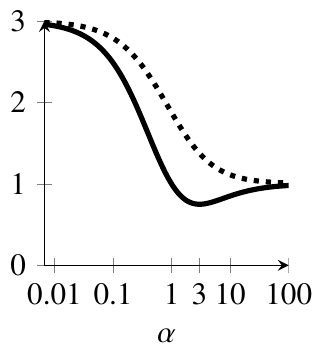}
\caption{$\Gamma\tau_{a}$ ($\Gamma\tau_c$) vs photon decay parameter $\alpha$ for an atom (cavity) in the solid (dashed) curve. }
  \label{F:3}
\end{figure}

In the waveguide-atom system, the effect of stimulated emission is also manifested in the lifetime of the atomic excitation.  We define such a lifetime as 
\begin{align}
\tau_a\equiv\frac{1}{2}\int_{0}^{\infty} dt \langle \sigma_z+1\rangle.
\end{align}
With this definition, a spontaneously decaying atom yields $\tau_a=1/\Gamma$, as expected.  In Fig. \ref{F:3}, we plot $\tau_{a}$ as a function of $\alpha$, which one may recall from Eq. \eqref{E:pulse} characterizes the temporal decay rate of the incident photon pulse. In the limit $\alpha>>1$, the atom lifetime is identical to the spontaneous emission lifetime $1/\Gamma$.  In this regime, the incident photon's duration is extremely short, which has no effect on the atom's lifetime.  In the limit $\alpha<<1$, $\tau_{a}=3/\Gamma$. In this regime, the atom decays spontaneously, and the photon interacts with a ground-state atom, exciting the atom, which then decays again. When $\alpha=3$, the lifetime of the atom is minimal and equals $\tau_{a}=0.75/\Gamma$. This corresponds to the condition for a maximum in the photon correlation term in Eq. \eqref{E:Prrll}, demonstrating that stimulated emission in this system is associated with a shortening of the atom lifetime.  In contrast, the lifetime of a cavity $\tau_c\equiv\int_{0}^{\infty} dt\langle a^{\dag}a\rangle$ is always longer than $1/\Gamma$, which it approaches from above as $\alpha \to \infty$.

 It is interesting to note that the maximum probability $P_{RR}=2/3$, achieved at $\alpha=2$, is equal to the maximum probability of cloning a quantum state in a universal quantum cloning machine \cite{Bruss1998,Lamas-Linares2002}.  However, the system here is not a universal quantum cloning machine.  If we inject a single photon in the even subspace, both of the outgoing photons will be in the even subspace.  Moreover, the initial single-photon wavefunction is not cloned in the outgoing two-photon wavefunction.  Despite this fact, we can consider the stimulated emission process studied here by using the language of photon cloning, if one is interested only in the outgoing direction of the photons and not the detailed photon wavefunction. Then, the photon cloning fidelity \cite{Lamas-Linares2002} maximizes at $\alpha=3$, yielding a fidelity of $\eta=P_{RR}+1/2P_{RL}=0.8125$.  The photon-number amplification factor is maximal at $\alpha=2$, with a value of $2P_{RR}=4/3$.  
 
In addition to the coupling between the atom and waveguide, the atom can also couple into nonguided modes, which are considered a loss mechanism.  In this case we define the $\beta$ factor as $\beta\equiv\frac{\Gamma}{\Gamma+\gamma_{ng}}$, where $\gamma_{ng}$ is the coupling rate into nonguided modes.  Large $\beta$ factors of ~0.9 and higher have been experimentally measured with the same geometry studied here \cite{Chang2006,Lund-Hansen2008,Thyrrestrup2010}.  In our analysis, a less-than-unity $\beta$ factor can be accounted for by making the replacement $\Omega\to \Omega-\mi\gamma_{ng}/2$ in the loss-free results.  For $\beta=0.9$, the results are modified only slightly, with the maximum of $P_{RR}$ now occurring at $\alpha\approx 2.04$, resulting in $P_{RR}\approx 0.63$.

In summary, we have provided a fully quantum-mechanical theoretical study of stimulated emission at its most basic level, involving an incident single-photon pulse and an atom in its excited state.  The study provides insights into this fundamental process of nature, including the role of photon pulse dynamics, and the photon correlation induced by the atom due to its inability to absorb more than one photon at a time.  Such a study may prove useful as one seeks to exploit such systems for manipulation of quantum states.

\appendix
\pdfbookmark[5]{Supplementary Information -- Calculation of $\phi_{eo}(x_c-t,x_d)$ in Eq. (4) and $\phi_{ee}(x_c-t,x_d)$ in Eq. (5) for the input state in Eq.(2) }{Supplementary Information -- Calculation of $\phi_{eo}(x_c-t,x_d)$ in Eq. (4) and $\phi_{ee}(x_c-t,x_d)$ in Eq. (5) for the input state in Eq.(2) }
\section{Supplementary Information -- Calculation of $\phi_{eo}(x_c-t,x_d)$ in Eq. (4) and $\phi_{ee}(x_c-t,x_d)$ in Eq. (5) for the input state in Eq.(2) }
\label{Supp}
In the $ee$ subspace, we have:
\begin{align}
&\lim_{t\to\infty}e^{-iHt}\ket{\text{in}}_{ee}=
\\&=\frac{1}{\sqrt{2}}\lim_{t\to\infty}\int\int dk_1dk_2e^{-iH_et}\ket{k_1k^+_2}\int dx\psi(x)\bra{k_2^+k_1}c^{\dag}_e(x)a^{\dag}\ket{0}.
\label{E:app1}
\end{align}
Above, we have explicitly written the limit $t\to\infty$, and inserted the resolution of the identity in the two-excitation manifold.   Using the results for $\ket{k_1^+,k_2^+}$ in Ref. \cite{Shen2007}, we have
\begin{align}
&\int dx\psi(x)\bra{k_2^+k_1}c^{\dag}_e(x)\sigma_+\ket{0}=\int dx\psi(x)\left(\bra{e_{k,p}}c^{\dag}_e(x)\sigma_+\ket{0}+\bra{e_{E}}c^{\dag}_e(x)\sigma_+\ket{0}\right),
\label{E:Ibra}
\end{align}
where 
\begin{align}
&\ket{e_{k,p}}=\frac{V}{2\pi\sqrt{\Delta^2+(\Gamma/2)^2}}\int dx \Bigg\{\Bigg[\left(\Delta-i\Gamma/2\right)\frac{e^{ik_1x}}{k_2-\Omega+\mi\frac{\Gamma}{2}}
\\&+\left(\Delta+i\Gamma/2\right)\frac{e^{ik_2x}}{k_1-\Omega+\mi\frac{\Gamma}{2}}\Bigg]\theta(-x)+\Bigg[\left(\Delta-i\Gamma/2\right)t_{k_2}\frac{e^{ik_2x}}{k_1-\Omega+\mi\frac{\Gamma}{2}}
\\&+\left(\Delta+i\Gamma/2\right)t_{k_1}\frac{e^{ik_1x}}{k_2-\Omega+\mi\frac{\Gamma}{2}}\Bigg]\theta(x)\Bigg\}c^{\dag}_e(x)\sigma_+\ket{0},
\end{align}
and
\begin{align}
\ket{e_E}=\sqrt{\frac{2\Gamma^2}{\pi}}\int dx\frac{e^{iEx/2-\Gamma |x|/2}}{E-2\Omega+2\mi\Gamma}c^{\dag}_e(x)a^{\dag}\ket{0},
\end{align}
are the parts of the eigenstate describing one excited atom and one photon.  Moreover, since we are interested in the limit $t\to\infty$, we may rewrite:  
\begin{align}
&\lim_{t\to\infty}\int dk_1dk_2e^{-iH_et}\ket{k_1k_2^+}=\lim_{t\to\infty}\int dk_1dk_2e^{-iEt}\ t_{k_1}t_{k_2}\ket{W_{k_1,k_2}}
\\&+\lim_{t\to\infty}\int dE\ e^{-iEt}t_E\ket{B_E}\label{E:Iket},\end{align}
Where $\ket{W_{k_1,k_2}}$ is the two-photon part of the propagating eigenstate, and $\ket{B_E}$ is the two-photon part of the bound eigenstate  as given in Ref. \cite{Shen2007}.  Additionally, $t_{k_{1,2}}=\frac{k_{1,2}-\Omega-\mi\Gamma/2}{k_{1,2}-\Omega+\mi\Gamma/2}$; $t_E=\frac{E-2\Omega-2\mi\Gamma}{E-2\Omega+2\mi\Gamma}$;  $E=k_1+k_2$ and $ \Delta=\frac{k_1-k_2}{2}$.
substituting the results in Eq. \eqref{E:Ibra} and Eq. \eqref{E:Iket} back into Eq. \eqref{E:app1} yields the $ee$ subspace out-state
\begin{align}
&\lim_{t\to\infty}e^{-iH_et}\ket{\text{in}}_{ee}=\frac{1}{\sqrt{2}}\lim_{t\to\infty}\int dk_1dk_2e^{-iEt}\ t_{k_1}t_{k_2}\ket{W_{k_1,k_2}}\times
\\&\int dx\psi(x)\bra{e_{k,p}}c^{\dag}_e(x)\sigma_+\ket{0}
\\&+\frac{1}{\sqrt{2}}\lim_{t\to\infty}\int dE\ e^{-iEt}t_E\ket{B_E}\int dx\psi(x)\bra{e_{E}}c^{\dag}_e(x)\sigma_+\ket{0}
\\&=\lim_{t\to\infty}\int dx_1dx_2\phi_{ee}(x_c-t,x_d)\ket{x_1,x_2}_{ee}\label{E:psiee},
\end{align}
where:
\begin{align}
&\phi_{ee}(x_c-t,x_d)=\Gamma\sqrt{\alpha}\theta(t-x_c-|x_d|/2)e^{2i\Omega(x_c-t)}e^{\frac{\Gamma}{2}(\alpha+1)(x_c-t)}e^{\Gamma(\alpha-1)|x_d|/4}\times
\\&\left\{1-\frac{\alpha+1}{2(\alpha-1)}\left[1-e^{-\Gamma(\alpha-1)|x_d|/2}\right]\right\}.
\end{align}
The out-state in the $eo$ subspace may be gotten from Eq. \eqref{E:psiouteo} with the help of the one-excitation eigenstate of the system in Ref. \cite{Shen2007}, and as a result we have: 
\begin{align}
\phi_{eo}(x_c-t,x_d)=\Gamma\sqrt{\alpha}\theta(t-x_c-|x_d|/2)e^{2i\Omega(x_c-t)}e^{\frac{\Gamma}{2}(\alpha+1)(x_c-t)}e^{\Gamma\left(\frac{1-\alpha}{4}\right)x_d}.\label{E:psieo}
\end{align}

 \bibliography{StimulatedEmission}

\end{document}